\documentstyle[psfig,twocolumn]{esapub}

\begin{document}

\setlength{\parindent}{0pt}
\setlength{\parskip}{ 10pt plus 1pt minus 1pt}
\setlength{\hoffset}{-1.5truecm}
\setlength{\textwidth}{ 17.1truecm }
\setlength{\columnsep}{1truecm }
\setlength{\columnseprule}{0pt}
\setlength{\headheight}{12pt}
\setlength{\headsep}{20pt}
\pagestyle{esapubheadings}

\title{\bf NGST AND THE PHYSICAL EXPLORATION OF GALAXY EVOLUTION
DURING THE ERA 2$<z<$5}
\author{{\bf Richard S.~Ellis} \vspace{2mm} \\
Institute of Astronomy, Madingley Road, Cambridge CB3 0HA, UK}

\maketitle
\begin{abstract}

NGST offers unprecedented opportunities for charting the `dark ages'
beyond the limits of the deep optical surveys conducted with HST. An
equally important motivation, however, is a detailed physical
understanding of the later stages of cosmic history including the
period 2$<z<$5 when most galaxies are thought to assemble and undergo
dramatic changes in their star formation rate, chemical content and,
ultimately, their morphological characteristics. An important question
we need to resolve is the extent to which established massive
systems might exist at moderate redshifts. In the context of recent
observational and theoretical progress in this area, I review the role
NGST could play in this redshift range taking into account the likely
progress made in the next decade with the new generation of 8-m 
class ground-based telescopes. 
\vspace {5pt} \\

Key~words: cosmology, galaxy evolution \& formation.

\end{abstract}

\section{`STAR WARS' - GROUND VERSUS SPACE IN 2007}
 
It is a time of rapid progress in observational cosmology and it seems
foolhardy to predict in detail what we might do scientifically in this
area with NGST in 2007. In this brief review I have preferred to take a
more strategic view of the scope and opportunities offered by NGST,
both in the context of expected progress with ground-based facilities
and with the goal of improving our knowledge of the physics of galaxy
formation in a redshift range where sources with stellar radiation
certainly exist. This contrasts with the exciting possibility of using
NGST to explore the `Dark Ages' before galaxies assembled where, at
present, only brave theorists can guide us (Loeb, Rees, this volume).

What we can say with certainty ten years from now is that there will be
an enormous increase in ground-based capability. At the last count we can
expect 13 $\times$ 6.5m-11m telescopes with a collective surface area
equivalent to over ten Keck telescopes. History has shown that we have
tended to {\it underestimate} rather than exaggerate the future impact
of our new facilities. An example in faint galaxy spectroscopy will
make this point clear. In the 1980's when many of us were campaigning
for new 8-m telescopes, 4-m telescopes were reaching what we considered
a hard limit of I=22, B=24 for absorption line galaxy spectroscopy.
Skeptics used to argue that an 8-m telescope would only push back this
frontier by at most a magnitude. In fact, Keck spectra of I=25-26, B=26-27
galaxies are routinely being gathered representing an order of magnitude
better performance than aperture scaling of old technology.  This reflects
the combination of many incremental advances in technology which appear
first on our newest telescopes (telescope performance, image quality,
detector characteristics, spectrograph throughput..) as well as personal
ambitions in an increasingly competitive area. One can hardly fail to
be impressed by the remarkable image quality achieved at an early stage
by the active primary on the ESO VLT (Giacconi 1998) which may herald
an exciting new era in high resolution ground-based imaging. For these
reasons, it seems sensible to consider rather carefully the improving
performance of these 8-m telescopes against that proposed for NGST.

\begin{figure}
\begin{center}
\leavevmode
\centerline{\psfig{file=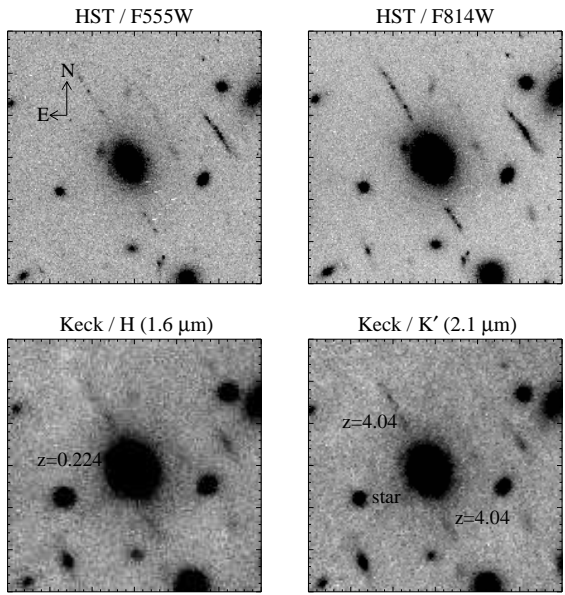,width=8.0cm}}
\end{center}
\caption{\em HST optical and Keck near-infrared imaging of distant
gravitationally-lensed arcs from the study of Bunker et al (1998)
demonstrating that ground-based telescopes are encroaching upon the
hitherto unique advantage of HST's image quality. Each panel
represents a 20 arcsec field.}
\label{fig:Figure1}
\end{figure}

We begin by examining the current symbiosis between HST and ground-based
telescopes. In my view the complementarity between HST and a large
ground-based telescopes has often been overstated. On the one hand HST
is competitive with larger ground-based telescopes for many cosmological
projects and, on the other hand, ground-based telescopes are 
increasingly encroaching into high resolution imaging territory that 
has traditionally been reserved for HST. Let us examine two remarks
that one frequently hears:

\begin{itemize}

\item {\it HST is a small-aperture telescope} - this is a misleading
comment for background-limited imaging at those wavelengths
(0.8$<\lambda<$1.8 microns) affected by airglow (which include those
increasingly important for cosmology) since the background in space at these
wavelengths is 40 times lower.  Roughly speaking HST has the capability
of a ground-based 6.5m telescope for this kind of work.

\item {\it HST offers unique spatial resolution compared to ground-based
telescopes} - this is certainly true in the UV/optical but recent Keck
near-infrared images (Bunker et al 1998, Figure 1) show that the gain can, on
occasions at least, be quite modest for many applications that have
traditionally been reserved for HST. Adaptive optics will, of course,
erode this gain.

\end{itemize}

Noting the likely improvement in ground-based technology, what then
could NGST uniquely provide in comparison to the impressive army of 8-m
telescopes we can expect to be in full swing by 2007? An interesting
comparison of the relative performance of Gemini and NGST has been
recently published by Gillett \& Mountain (1998). Whilst a number of
assumptions have to be made in such comparisons, two are worth bearing in
mind in understanding the difference between their conclusions and those
by the NGST study team. First, Gillett \& Mountain assumed Adaptive
Optics on Gemini would routinely deliver near diffraction-limited
performance at wavelengths longer than 1 micron (a Strehl of 0.8 at
K). Without AO, Gemini would deliver about 0.4 arcsec at K (matching
that of Figure 1). Secondly, they assumed that NGST would be limited to
fairly short exposure times because of variable cosmic ray hits. Looking
at their plots (reproduced in part by permission of the authors in Figure
2) there are two regimes of importance:

\begin{figure*}[!ht]
\begin{center}
\leavevmode
\centerline{\psfig{file=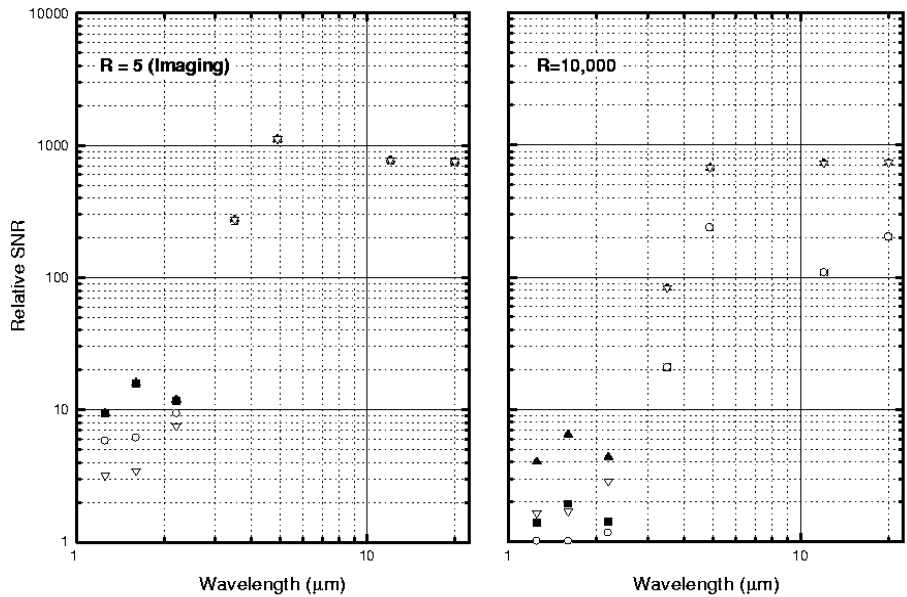,width=12cm}}
\end{center}
\caption{\em A comparison of the performance of NGST and an adaptive
IR-optimised ground-based 8-m telescope reproduced from the study of
Gillett \& Mountain (1998). The panels show the relative point source
signal/noise ratio for R=5(imaging) and R=10,000(spectroscopy). See
original article for symbol explanations, assumptions and caveats.  The
principal gain of NGST lies in the wavelength range $\lambda>$3$\mu$m.
Within the 1-2$\mu$m range, there appears to be a significant gain only
for broad-band imaging.}
\label{fig:Figure2}
\end{figure*} 

\begin{itemize}

\item{0.5$<\lambda<$2.2 microns:} here the gain of NGST is claimed to
be only appreciable for broad-band imaging. Such gains would be reduced
if ground-based telescopes could find an efficient way to perform wide
field OH-suppressed imaging, which seems possible given recent
successes in OH-suppressed spectroscopy
(http://www.ast.cam.ac.uk/~optics/cohsi/cohsi.htm). For high dispersion
spectroscopy, the gain of NGST is claimed to be quite modest, but this
depends in detail on the projected performance of infrared detectors
and several of the other assumptions made.

\item{$\lambda>$2.5 microns:} here the gains become substantial 
($\simeq$10$^3$) for all likely observing modes. This wavelength 
region is strategically important for many reasons but especially
spectroscopy of redshifted diagnostic lines like H$\alpha$
beyond $z$=2.5 and for the analysis of continuum radiation from older 
stellar populations to $z\simeq$5; optical/UV radiation from an early 
generation of young stars could also in principle be seen to very 
high redshifts.

\end{itemize}

In summary, NGSTs gain over ground-based telescopes is greatest beyond
2.5 microns and hence, very effective in exploring the range 2$<z<$5.
Before examining the techniques we might use to explore this region, it
is helpful to understand some of the controversies surrounding our
present understanding of how galaxies form and evolve, noting both
observational and theoretical progress.

\section{PROGRESS IN GALAXY FORMATION}

The now familiar way to examine recent progress in galaxy formation and
evolution is via the comoving volume-averaged star formation history
that has been derived observationally from the optical/IR census of
galaxy luminosities selected in various ways (Madau 1997) and predicted
theoretically from hierarchical cosmologies where gas cooling around
cold dark matter (CDM) halos is inhibited by various feedback processes
(Cole et al 1994, Kauffmann et al 1994, Baugh et al 1998). Although the
redshift dependence of the star formation history remains
quantitatively controversial (Blain et al 1998, Hughes et al 1998) and
its interpretation is not uniquely tied to hierarchical theories of
structure formation, such caveats will not be important in the
following discussion. Rather than argue about the detailed form of the
star formation history, I want to concentrate on more general
implications supported by both theory and observations.

Foremost, what we have learnt from both theoretical and observational
studies of the star formation history is that galaxy formation is an {\em
extended process}, rather than the single event once imagined. But is
there still room for a population of objects that collapsed monolithically
at some redshift, either at beyond the current observational redshift
window or, within it, perhaps shrouded in dust? The answer to this
question depends on the stellar mass of high redshift star-forming systems
which is currently very poorly constrained. 

Extended galaxy formation is a central assumption in hierarchical
theories because of the interplay between feedback and gas cooling
so it remains observationally very important to explore the physical
properties of systems at high redshift. Indeed, it is worth highlighting
the remarkable contrast between traditional and CDM viewpoints for the
formation of giant elliptical galaxies.  As ellipticals are compact
objects showing little rotation, the traditional viewpoint asserts these
form via monolithic dissipationless collapse at very high z. In contrast,
hierarchical cosmologists believe in the slow assembly of disk systems
around dark matter halos and these later merge to form ellipticals.
The question of whether galaxies assembled hierarchically over a large
range in redshift (with perhaps a peak of activity at $z\simeq$1-2) or
collapsed monolithically clearly has a profound impact on the likelihood
or otherwise of finding high redshift stellar populations (as opposed
to only gas clouds).

So far, observational evidence has not convincingly come down in favour
of either hypothesis. On the one hand, examples are found of high
redshift radio galaxies which appear genuinely old (Dunlop 1998) and
luminous cluster galaxy populations at intermediate redshifts display
homogeneous ultraviolet-optical colours (Ellis et al 1997, Stanford et al 
1997). Both studies indicate significant star formation must have
occurred before $z\simeq$3. But, as Kauffmann (1995) and Governato et al 
(1998) have pointed out, in biased models of galaxy formation we can 
expect accelerated evolution in dense regions. Studies of rare objects 
at high z do not provide constraints which can be applied with confidence
to the wider population.

\begin{figure} 
\begin{center}
\leavevmode
\centerline{\psfig{file=Figure3a.ps,width=7.5cm}}
\vspace{0.5truecm}
\centerline{\psfig{file=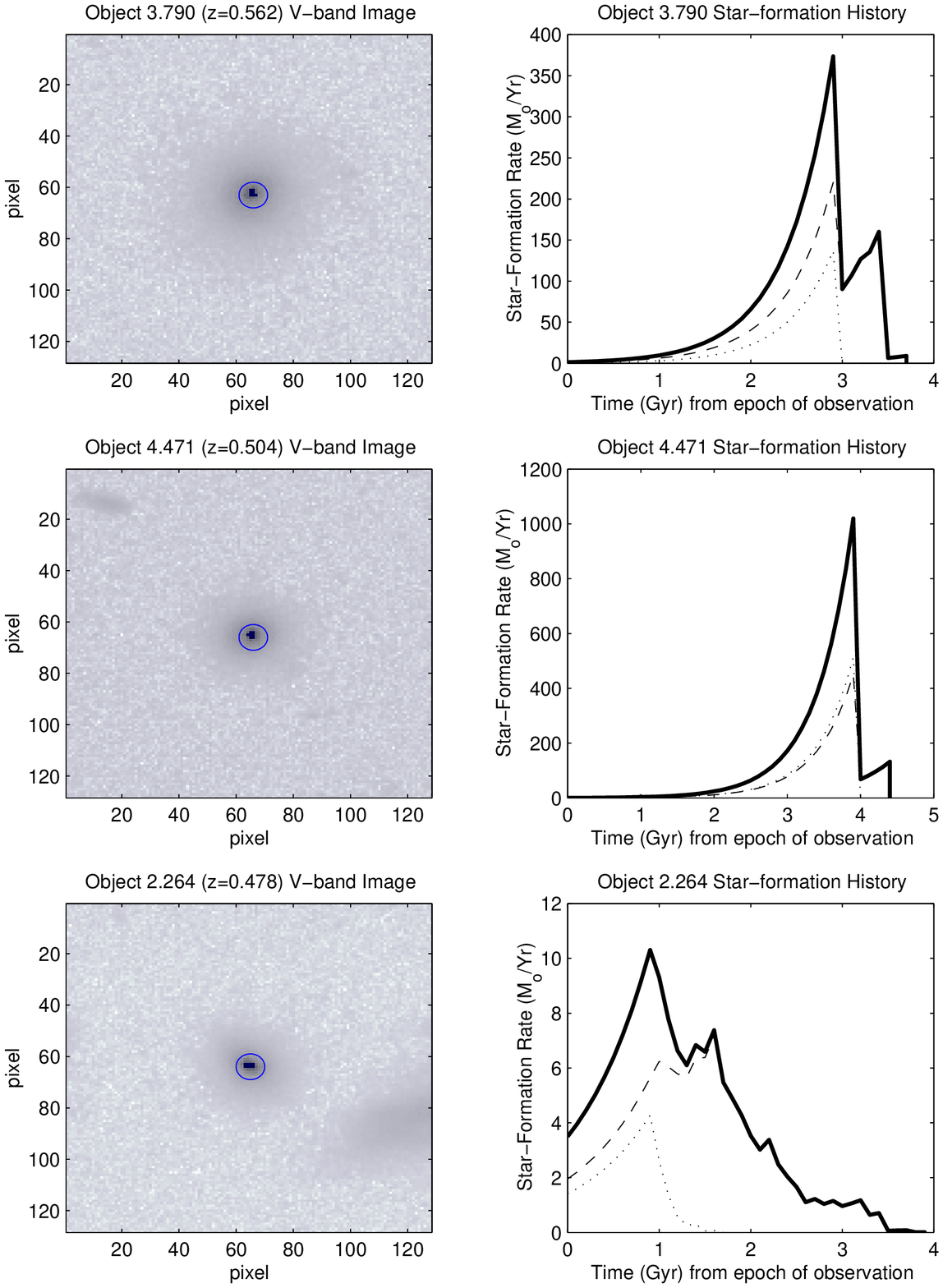,width=7.5cm}}
\end{center}
\caption{\em (a) A montage of field ellipticals selected
morphologically from the joint HST-infrared imaging database of
Menanteau et al (1998). Such samples demonstrate a paucity of red
systems compared to expectations for models where ellipticals completed
their star formation prior to $z\simeq$3 (in contrast to their
clustered counterparts).  (b) Internal pixel-by-pixel population
studies for HDF galaxies of known redshift can be used to infer the
prior star formation history of morphologically-selected sources
(Abraham et al 1998). A significant proportion of field ellipticals
(e.g. bottom panel) reveal colour inhomogeneities providing further
evidence for recent activity}
\label{fig:Figure3}
\end{figure} 

A crucial goal is the evolutionary history of field ellipticals. Do we
see a depletion of this population at modest redshift as expected in
hierarchical models? Kauffmann et al (1996) claimed to find a shortage
of red (V-I) objects in the CFRS redshift survey which, they claim, is
consistent with the continued production rate of ellipticals from
mergers in CDM models (see also Zepf 1997). However, colour alone is a
very poor guide to morphology (Schade et al 1998) and utilising HST
images in conjunction with deep infrared photometry, Menanteau et al
(1998) have conducted a new extensive search based on HST morphology.
They, similarly, find a shortage of passively evolving spheroidals to 
K=20.  Moreover, in a new approach, Abraham et al (1998) have analysed 
the internal pixel-by-pixel colour distribution of ellipticals 
of known redshift in the HDF and find a sizeable proportion show 
a diversity of internal colours suggesting further evidence of recent 
activity (Figure 3).

In summary, the present evidence for established systems at high 
redshift includes:

\begin{enumerate}

\item{} The presence of radio galaxies with red continua and other sources
with prodigious star formation rates at high redshift including the
recently-discovered population of sub-mm sources (Blain et al 1998) and
even the Lyman break galaxies, examples of which can be found with star 
formation rates approaching 100 M$_{\odot}$ yr$^{-1}$ (Steidel et al 1998).

\item{} Kinematic data which suggest damped Lyman alpha systems are
well-established thick, rapidly rotating disks (Prochaska \& Wolfe 1997,
1998). Alternative interpretations of this data have been presented,
however, which are consistent with protogalactic clumps undergoing
infall within dark matter halos (Haehnelt et al 1997).

\item {} The abundance of well-formed spirals at z=1 (Lilly et al 1998)
whose stellar populations are consistent with a declining activity
since at least a redshift 1.5-2 (Abraham et al 1998). This seems to be
in conflict with the later formation of stellar disks necessary in
hierarchical theories in order to avoid a serious reduction in the
angular momentum by tidal redistribution (Weil et al 1998).

\end{enumerate}

Evidence which suggests the bulk of the population formed late include: 

\begin{enumerate}

\item{} The absence of high redshift sources in K-selected samples.
The `K-band redshift survey test' was originally proposed by
Broadhurst et al (1992) and has been revisited in the framework of CDM
models and the data of Cowie et al (1996) by Kauffmann \& Charlot
(1998).  This shortfall of luminous K-band objects at high redshifts is
consistent with the marked change in slope of the near-infrared counts
(Ellis 1997) now confirmed to very faint H limits with NICMOS (Yan et
al 1998)

\item{} The small angular sizes of HST images in the HDF and in other deep
fields (Pascarelle et al 1997) which are highly suggestive of sub-units
which later merge to form normal galaxies. Bouwens et al (1998) have
simulated the appearance of a $I<$22 sample of galaxies of known redshift
when placed at greater distance in the HDF, allowing carefully for instrumental
and surface brightness dimming, and concluded there has been strong size
and number density evolution since $z\simeq$3.

\item{} The remarkably rapid and recent decline in galaxies with
irregular morphology (Glazebrook et al 1995, Brinchmann et al 1998).
The fate of these systems remains unclear but their delayed
contribution to the luminosity evolution of the Universe gives strong
support to the suggestion that star formation on galactic scales may be
governed by feedback processes in addition to simple gravitational
collapse of gas clouds around dark matter halos (Babul \& Rees 1992).

\end{enumerate}

Much of the evidence for massive systems at high redshift therefore
rests on extreme objects. A valuable test of whether this is simply
accelerated evolution due to bias is to examine the spatial
distribution as a function of their number density. If these
luminous objects are unrepresentative of the history of less massive
galaxies, as CDM proponents claim, then one expects to find a high bias
for the rarer galaxies. Steidel et al (1998) claim to detect the first
tentative evidence of such a density-dependent bias in a correlation
analysis of Lyman break galaxies selected in various ways. This serves
to highlight the important connection between galaxy evolution and
the large scale distribution of faint sources.

Concerning the presence or otherwise of primaeval galaxies, a topical
question at the time of writing is the nature and redshift distribution
of the population of sub-mm sources being found with the SCUBA array on
the James Clerk Maxwell Telescope (Hughes et al 1998, Barger et al
1998). The negative k-correction for dusty sources observed at 850 $\mu$m 
implies luminous sources such as Arp 220 could be detected to
$z\simeq$10. However, where optical counterparts can be checked, the
indication is that the SCUBA sources are coeval with the more 
modestly star-forming galaxies selected in the UV/optical (Smail et al 
1998). However, even if more extensive redshift identifications confirm
this is the case, an important feature of the early SCUBA results is 
the rapidity with which the far-IR background has been resolved into 
sources suggesting most of the high z dust emission is confined to a 
population of sources whose comoving volume density is nowhere near as 
high as that required to make up the UV background. Does this, in turn, 
imply that the SCUBA sources represent a fundamentally different population
from the luminous high $z$ galaxies being found at optical wavelengths?

\section{A PHYSICAL UNDERSTANDING OF GALAXY FORMATION}

It is clear from the above discussions that even a complete derivation
of the star formation history of the Universe from various diagnostic
measurements will only represent the most basic step forward in
understanding galaxy formation. Such a global measurement integrates
over all luminosities and types and therefore hides most of the
important physical details. Moreover, the apparent agreement between
observation and theory can hardly represent a robust test of CDM, at
least when it concerns testing ingredients such as feedback, merging
and morphological evolution. We really need to test the more basic
physical principles of any model and that means securing detailed
properties of individual objects. Ultimately we wish to understand the
dynamical state of forming systems and crucially their masses.

This might seem something of a tall order but NGST's superior performance at 
2.5$<\lambda<$5 microns can help. Examples include:

\begin{itemize}

\item {} Imaging at longer wavelengths than is possible with ground-based
facilities. This allows us to trace established stellar populations at high
redshift and hence to measure stellar luminosities less affected by
dust and young stars. 

\item {} Detailed internal physical properties will become available through
resolved, or integral field, spectroscopy including dynamical
characteristics, excitation properties, dust content and star formation
rates. Other articles in this volume will develop the theme of integral field
spectroscopy more fully. Here I will point out some of the issues we face.

\end{itemize}

The rest-frame near-infrared luminosity can
be tracked with diffraction-limited imaging to $z\simeq$5 or so
yielding an integrated estimate of the established stellar mass.  A
major advantage of working in the near-infrared is that the K-band luminosity
of an evolving system seems to be largely independent of its previous
history (Kauffmann \& Charlot 1998). However, on shorter timescales
($<$1 Gyr) there could be biases arising from transient populations.
For example, red supergiants and AGB stars may temporarily raise the
visibility of star-forming galaxies in the K-band. Calibrating these
effects will require a more complete understanding of stellar
populations at infrared wavelengths (Charlot et al 1996).

\begin{figure}[h] 
\begin{center}
\leavevmode
\centerline{\psfig{file=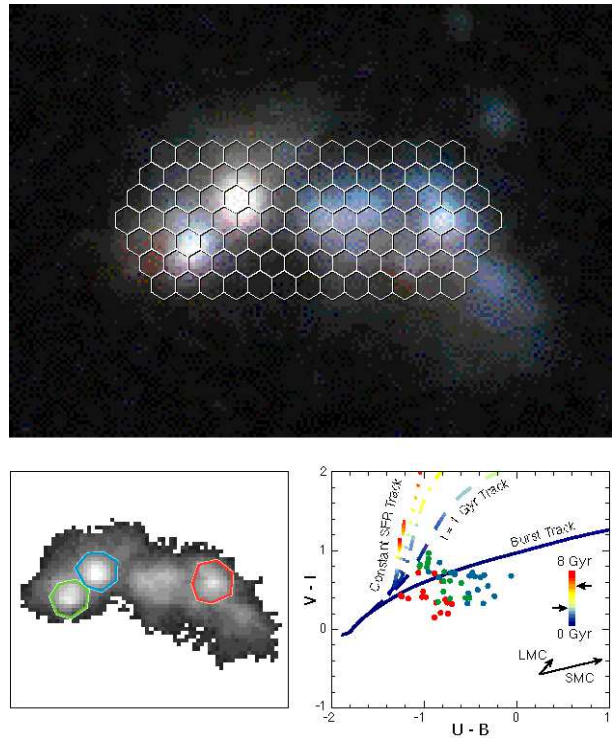,width=8cm}}
\end{center}
\caption{\em Integral field spectroscopy and multicolour HST imaging 
represent a powerful combination for unravelling the physical properties
of a distant irregular. Upper panel shows the HDF image of a $z$=1.355
galaxy with irregular morphology which is well-suited for integral
field spectroscopy (each fibre segment represents a diameter of 0.15 arcsec
or $\simeq$600h$^{-1}$pc). The lower panel analyses the pixel-by-pixel
colour distribution for physically-distinct components (as indicated
by each selected region) in the context of different star formation
histories viewed at different times (solid and dashed curves). Dynamical
data will be essential to resolve the question of whether such a system
has recently merged or is established and undergoing stochastic bursts
of activity.}
\label{fig:Figure4}
\end{figure}

The motivation for IFU spectroscopy is easy to understand.  Is a
distant irregular (c.f. Figure 4) a well-established dynamical entity
undergoing sporadic star formation (Noguchi 1998) or are we witnessing
the arrival of physically-distinct sub-components with a chaotic
velocity field?  At the 4-m telescope level, IFU spectroscopy
has proved to be extremely demanding in telescope time; only luminous
or gravitationally-lensed sources have yielded interesting results
(Soucail et al 1998). The problem is that there is a huge dynamical
range in surface brightness in any object and this will be exascerbated
at high redshift. Whilst it is tempting to restrict spectral analyses
to those bright spots of star formation which conveniently provide
intense emission lines, local studies already illustrate that line 
widths give only a lower limit on the circular velocities 
(Lehnert \& Heckman 1996, Rix et al 1997).  We really need to 
have good sensitivity at much larger radii.

\section{SUMMARY}

\begin{enumerate}

\item {} The design and scientific strategy of NGST should take into account
the dramatic progress we can expect from ground-based 8-10m telescopes
in the coming decade. In particular, it will be important to scrutinise
carefully the achievements of the IR- optimised 8-m telescopes in the JHK
windows.

\item{} Nonetheless, the clear area where NGST will excel lies longward of 2
microns and this offers the opportunity of studying the assembling
galaxy population in the redshift range 2.5$<z<$5.

\item {} A key goal for the future is the determination of stellar and total
masses for representative subsets of the evolving population. This will
necessitate mid-IR images and high resolution IFU spectroscopy to very
low surface brightness limits in order to yield estimages of the
integrated stellar mass and the internal dynamics.

\end{enumerate}

\section*{ACKNOWLEDGEMENTS} 

I acknowledge numerous discussions with my colleagues at Cambridge and
collaborators on the CFRS/LDSS redshift survey and HDF programmes and 
thank Bob Abraham, Alfonso Aragon, Jarle Brinchmann, Roger Davies, Fred
Gillett, Matt Mountain, Martin Rees and Chuck Steidel for helpful input.

\end{document}